\documentclass[12pt]{article}
\begin{document}
\begin{center}
{ \large WHAT THE $K_{e4}$ DECAY TELLS US ON A VALUE OF $\sigma$-PARTICLE MASS AND ON A NATURE OF THE SPIN-1 MESONS}  \\
{\bf E. Shabalin   \footnote{E-mail address: shabalin@itep.ru} }\\
 { Institute for Theoretical and Experimental  Physics , Moscow, Russia}
\end{center}
\begin{abstract}
The data on the form factors of the $K_{e4}$ decay permit to  fix a value of the  non-specified in the theory parameter entering  in  the lagrangian of the system of $0^+$ and  $0^-$   mesons . Then, for a mass of the lightest  $\sigma$-meson we find:
$m_{\sigma}$=663 MeV.
As for a nature of the spin-1 mesons, also contributing into $K_{e4}$ form factors, the data do not allow to interpret them as
the gauge bosons of the chiral theory.
\end{abstract}

\section{Introduction}
In present paper we argue in favour of the chiral theory incorporating not only the  pseudoscalar mesons, but also the scalar ones.
A reason is that appearing as the intermediate particles in the pions interaction, the last allow to understand and explain the
dependence of the proceses with pions of energy.  Besides, the low-energy scalar resonances exist and their specific properties
get  an explanation in such a theory.

The data on the form factors of the $K_{e4}$ decay obtained in
\cite{1}, where 400,000 events of the decay $K^+ \to \pi^+\pi^- e^+\nu$
had been observed, will give us  a possibility  to explain the role of the scalar mesons in this decay and, in addition, to get an
answer on  a nature of the spin-1 mesons, also contributing into $K_{e4}$ decay.

In QCD, the spinless flavoured objects are $\bar q_Rt^aq_l$, and
their Hermitian conjugate. These objects incorporate the parts
with the opposite parity. To reproduce this property in the
lagrangian of the real particles, it must be written in terms of
the matrix
\begin{equation}
  U=(\sigma_a + i \pi_a)t_a
\label{1}
\end{equation}
where $t_0=\sqrt{\frac{1}{3}}I, t_{1,...8}=\sqrt{\frac{1}{2}}\lambda_{1,...8}$,  and
$\sigma_a$ , $\pi_a$ are the nonets of the scalar and the  pseudoscalar mesons.
This idea is not original and there had been a number of interesting models \cite{2},
\cite{3}, but we pass right away to the final form of the lagrangian  for the spinless fields containing all forms of breakdown of chiral symmetry \cite{4}, \cite{5}:
\begin{equation}
\begin{array}{l}
L^{meson}=\frac{1}{2}Tr(\partial_{\mu}U\partial_{\mu}U^+)- cTr(UU^+ -A^2t^2_0)^2 - c\xi(Tr(UU^+-A^2t^2_0))^2 +  \\
  \frac{F_{\pi}}{2\sqrt{2}}Tr\{M (U+U^+)\} +   \frac{\Delta m^2_P F^2_{\pi}}{12} \cdot\{\frac{i}{2}Tr(\ln\frac{U}{U^+})\}^2.
\label{2}
\end{array}
\end{equation}
Here the parameter A characterises a strength of spontaneous  breakdown of the symmetry between the scalars and pseudoscalars,
the  matrix $M$ that in the limit  $m_u=m_d$ has the form
\begin{equation}
M= \sqrt{1/3}(2m^2_K+m^2_{\pi})t_0-2\sqrt{2/3}(m^2_K-m^2_{\pi})t_8
\label{3} \end{equation}
produces the hard  breaking $SU(3)$ symmetry and the last term and the term proportional to
the parameter $\xi$ lower the $U(3)_L\otimes U(3)_R$ symmetry down to
 $SU(3)_L\otimes SU(3)_R$  symmetry  due to mixing of
 $\pi_0$  and $\sigma_0$ state with the glueball states $\left(\frac{\alpha_s }{\pi}G^a_{\mu \nu}\tilde G^a_{\mu\nu}\right)$ and
$\left(\frac{\alpha_s}{\pi}G^a_{\mu\nu}G^a_{\mu\nu}\right)$ respectively.
The values of the constants $\Delta m_P$ and $\xi$ are not specified  in the theory;
they can be determined by fitting the theoretical predictions to the experimental data.
As for the masses of the scalar mesons, one finds from (2) :
\begin{equation}
m^2_{\sigma_{\pi}}-\mu^2=(m^2_K-\mu^2)[(R-1)(2R-1)]^{-1}\equiv \Lambda^2,
\label{4}
\end{equation}
\begin{equation}
m^2_{\sigma_K}-\mu^2=\Lambda^2(2R-1)R
\label{5}
\end{equation}
where $\mu^2=m^2_{\pi}$ and
\begin{equation}
R=F_{K}/F_{\pi}
\label{6}
\end{equation}
The parameter $R$ can be determined from $K, \pi \to \mu\nu$ decays, or  from the
relation (4) where the isotriplet scalar meson $\sigma_{\pi}$ is  identified with the
resonance $a_0(980)$. In the last case  we find for the mean value of $R$ :
\begin{equation}
R=1.176.
\label{7}
\end{equation}
This value is a little smaller (by 1 persent) than the one following from the $K, \pi \to\mu\nu$ decays.  But we shall use just the magnitude (7) for the self-consistency
of the further results.

For the isosinglet scalar  mesons, the mass formulae are more complicated than those
in (4), (5), since $\sigma_0$ is mixed wth the scalar gluonium as well as with $\sigma_8$  owing to breakdown of $SU(3)$ symmetry. As a result, the physical
isosinglet scalar mesons are
\begin{equation}
\begin{array}{l}
\sigma_{\eta'}=\sigma_0 \cos\theta_S + \sigma_8 \sin\theta_S    \\
\sigma_{\eta}=-\sigma_0 \sin\theta_S + \sigma_8\cos\theta_S.
\label{8}
\end{array}
\end{equation}
For these states we obtain:
\begin{equation}
\begin{array}{l}
m^2_{\sigma_{\eta'}}-\mu^2=\Lambda^2 \{1+2R(R-1)(\cos\theta _S - \sqrt{2}\sin \theta_S)^2  +  \\
\frac{1}{3}\xi[(2R+1)\cos\theta_S - 2\sqrt{2}(R-1)\sin\theta_S]^2 \} ,  \\
m^2_{\sigma_{\eta}} - \mu^2=\Lambda^2\{1+2R(R-1)(\sin\theta_S+ \sqrt{2}\cos
\theta_S)^2 +  \\
\frac{1}{3}\xi  [(2R+1)\sin\theta_S + 2\sqrt{2}(R-1)\cos\theta_S]^2\}.
\label{9}
\end{array}
\end{equation}
The coupling constants of $\sigma_{\eta'}$ and $\sigma_{\eta} $ to the pseudoscalar
mesons also depend  on  the parameter $\xi$. The expressions for them are adduced
 in Appendix. The mixing angle $\theta_S$ is connected with the parameters $R$  and
$\xi$ by the relation
\begin{equation}
 \theta_S= \frac{1}{2} \arctan\{2\sqrt{2} \frac{1+\xi (2R+1)(3R)^{-1}}{1-\xi(2R+1)^2
[6R(R-1)]^{-1}[1-8(R-1)^2(2R+1)^{-2}]}\}
\label{10}
\end{equation}

As it follows from  (9) and  Appendix, at $\xi=0$  only the one $\sigma_{\eta'}$ interacts
with two pions and its mass coincides with a mass of the $\sigma_{\pi}$ meson  that
is identified with the resonance  $a_0 $  with the mass 980 MeV.  But the numerous
 searches for $\sigma$ meson have
shown that a mass of the lightest $\sigma$ meson has to be smaller
than 700 MeV. \cite{6}. In our theory the last condition can be
fulfilled, if the parameter $\xi$ is negative, but $|\xi|\le
\frac{1}{3}$.This limitation is connected with a requirement of
positiveness of energy in the theory. As the parameter $\xi$
enters into the amplitudes originated by isos.inglet  $\sigma$
mesons, to extract its magnitude from the experimental data, we
need to pick out just  this part from the  data. In particular, to
exclude from the data  a contribution originated by the spin-1
mesons. However, a value of their contribution into considered
here processes depends on their nature.

For a long time, the spin-1 mesons were interpreted as the gauge  bosons of  the local
chiral symmetry \cite{7}.   But an availability of a mass, does not permit to be sure
 that the successful predictions being fair in such a theory in limit $m_{V,A}=0$,
remain to be fair at $m_{V,A}\ne 0$ \cite{8} . This idea led to reconsideration
of a nature of the spin-1 mesons \cite{8}- \cite{12}.

In \cite{9}, it was shown that a treatment  of the spin-1 mesons as the divergences
 of the corresponding antisymmetric tensor fields permits to consider them, on  a level
with the spinless mesons, as the members of the $(3, \bar3)+(\bar 3, 3)$
representation of the $S U(3)\otimes SU(3)$  chiral group.

 In \cite{10}, \cite{11}, it was stressed that "there is no
proof for the existence of dynamical gauge bosons of local chiral
symmetry in QCD" and "there is nothing  special about vector and
axial-vector mesons compared  to scalar, pseudoscalar or any other
meson resonances".

In the paper \cite{12}, the idea that  spin-1 mesons are the gauge bosons was
rejected proceeding from a principle of necessity of one-to-one  coincidence of the
general properties between the quark constructions in the quark-gluon space and
the physical particles in real world. The quark combinations
\begin{equation}
\bar q\gamma_{\mu} t_aq,    \qquad  \bar q\gamma_{\mu}\gamma_5t_a  q
\end{equation}
 associated in  the gauge theory
with the spin-1 particles, do not satisfy the subsidiary  condition
 $\partial_{\mu}V_{\mu}(x)=0$ and $\partial_{\mu}A_{\mu}(x)=0$ eliminating
the scalar components in the real spin-1 objects.
 Interpretation of the spin-1 fields
as the divergences of the antisymmetric tensor fields liberates one from a necessity
to  put on the subsidiary condition "by hands".
However, the examples of the evident discrepancy between the values of the observed
effects and their values calculated in the framework of gauge theory , were not presented in \cite{8}-\cite{12}. Such an example will be considered here.

\section{The amplitude of the $K^+ \to \pi^+\pi^-  e^+ \nu $   decay}
The matrix element of this decay has the following form:
\begin{equation}
\begin{array}{l}
M=(2\pi)^4 \delta^4(k-p-p'-p_e -p_{\nu})(8E_K E_{\pi}E_{\pi'})^{-1/2}\times    \\
 (G_FV_{us}/\sqrt{2}m_K) \times   \\
< \pi^+(p)\pi^-(p')|A_{\mu}^{(K)} +V_{\mu}^{(K)}|K^+(k)>\times         \\
   \bar\nu(p_{\nu})\gamma_{\mu}(1+\gamma_5)e(p_e).
\end{array}
\label{12}
\end{equation}
where $m_K$ is a mass of $K$ meson and
\begin{equation}
\begin{array}{l}
<\pi^+(p) \pi ^-(p')|A_{\mu}^{(K)}|K^+(k)>=f_1(p+p')_{\mu}+f_2(p-p')_{\mu} +\\
f_3(k-p-p')_{\mu},
\end{array}
\label{13}
\end{equation}
\begin{equation}
<\pi^+(p)\pi^-(p')|V_{\mu}^{(K)}|K^+(k)>=
  \frac{if_4}{m^2_K} \epsilon_{\mu\nu\sigma\tau}k_{\nu}p_{\sigma}p'_{\tau}.
\label{14}
\end{equation}
A probability of the $K_{e4}$ decay practically completely is determined by contribution of the terms proportional to $f_1$ and $f_2$ in (13).The part  proportional  to     $f_3$ is very small  because of proportionality to $m_e/m_K$.
The part  proportional to   $f_4$ gives a contribution of order of $0.04\%$, and
can be neglected in view of $2\%$ uncertainty of the measured up to now
 probability itself. So that, we shall  consider in  the present paper the  structure
of $f_1$ and $f_2$ only.

The current algebra methods  and  PCAC allow to evaluate the magnitudes of $f_1$
and  $f_2$ in the soft-pion limit, that is, at $(p+p')^2=\mu^2$ \cite{13}. More simply,
their values  can be found in a theory with the non-linear realization of chiral
 symmetry \cite{14}. At $(p+p')^2=\mu^2$:
\begin{equation}
f_1=f_2= -m_K/(\sqrt{2}F_{\pi}), \qquad  F_{\pi}=93\: \mbox{MeV}
\label{15}
\end{equation}
At the constant  $f_{1,2}$, a probability of the $K^+ \to \pi^+\pi^- e^+\nu$ decay
is \cite{15}
\begin{equation}
w^{th}=\frac{G^2_F V^2_{us}m^5_K}{2^{14}3 \pi^5}(f^2_1\cdot3.923\cdot
10^{-3}+f^2_2\cdot 7.728 \cdot 10^{-4}).
\label{16}
\end{equation}
At the magnitudes (15)
\begin{equation}
w^{th}=1.357\cdot 10^3 s^{-1}
\label{17}
\end{equation}
This value is more than two times smaller of the experimental one:
\begin{equation}
w^{exp}=3.302 (1\pm 0.022)\cdot 10^3 s^{-1}
\label{18}
\end{equation}
It means that a dependence of  $Q^2$ ,   at  least  $f_1$ , is
significant . But the experimental data show that  in  the region
$4\mu^2\le Q^2\le m^2_K$ the form factors change very slowly.
Consequently, a growth of the form factors must occur in the
non-physical region $\mu^2\le Q^2\le 4\mu^2$. A fast increase of
$f_1( Q^2)$ is possible, if the pair $\pi\pi$ appears from decay
of the intermediate $\sigma$ meson \cite{5}. Then a presence of
the structure $(m^2_{\sigma}-Q^2)^{-1}$ in the expression of
$f_1(Q^2)$ can ensure a desirable magnitude of $f_1(Q^2)=4\mu^2$
at the corresponding value of $m_{\sigma}$. In the point
$Q^2=4\mu^2$  a fast growth $f_1$ converts into  very slow
increase at $Q^2>4\mu^2$. The mechanisms originating this effect
are the following. The first - an appearance of the imaginary part
in the propagator of $\sigma$ meson \cite{5}. The second - an
appearance  of the form factor in the vertex $\sigma\pi\pi$
decreasing its value with a growth of $Q^2$ \cite{12}. And the
last - a rescattering of the final pions \cite{16}. The detailed
study of these effects is supposed to present elsewhere.  Such a
study is particularly desirable in view of the observed in
\cite{1} deviation of the form factor $f_1(Q^2)$ from the expected
behavior. At  present, a value  of mass of the $\sigma$ meson,
identified with the resonance $f_0(500)$, is very  uncertain.
\cite{17}. Our approach will permit to fix a value of $m_{\sigma}$
with a good accuracy.

\section{A contribution of the intermediate scalar mesons into  form factors
$f_{1,2}$.}
In our theory  the axial currents are given by
\begin{equation}
A_{\mu}^i=d_{ijk}(\sigma_j\partial_{\mu}\pi_k - \pi_j \partial_{\mu} \sigma_{k})
\label{19}
\end{equation}
where the $d_{ijk}$ coefficients are those defined by Gell-Mann
and $d_{0jk}=\sqrt{2/3}\delta_{jk}$. The states $\sigma_0$ and
$\sigma_8$ have the non-zero expectation  values:
\begin{equation}
<\sigma_0>=\frac{F_{\pi}}{\sqrt{6}}(2R+1), \qquad <\sigma_8>=-\frac{2F_{\pi}}
{\sqrt{3}}(R-1).
\label{20}
\end{equation}
In the terms of the physical states (8), the axial strange current
looks as the following:
\begin{equation}
\begin{array}{l}
A^{(k)}_{\mu}=\frac{i}{\sqrt{6}}\{[2\sqrt{2}\cos\theta_S - \sin\theta_S](\sigma_{\eta'}K^- ) -
[2\sqrt{2}\sin\theta_S +\cos\theta_S](\sigma_{\eta} K^- ) \}\cdot    \\ (p_{\sigma}+p_K)_{\mu} -
i\sigma_{\bar K^0} \pi^-  \cdot (p_{\sigma_K} + p_{\pi})_{\mu}.
\end{array}
\label{21}
\end{equation}
A contribution of $\sigma$ mesons into the amplitude (13) is represented by the matrix elements
$<\pi^+ \pi ^-|\sigma><\sigma|A_{\mu}^{(K)}|K^+>$ and $<\pi ^-|A_{\mu}^{(K)}|\sigma_{K^0}><\sigma_{K^0}|\pi^+ K^->$.
So that,
\begin{equation}
\begin{array}{l}
<\pi^+(p)\pi^-(p')|A^{(K)}_{\mu}|K^+(k)>=\frac{1}{\sqrt{6}} \{\frac{2\sqrt{2}
\cos\theta_S - \sin\theta_S}{m^2_{\sigma_{\eta'}} -  (p+p')^2}\cdot  g_{\sigma_{\eta'}\pi\pi} -      \\
\frac{2\sqrt{2}\sin\theta_S + \cos\theta_S}
{m^2_{\sigma _{\eta}}  - (p+p')^2)}\cdot g_{\sigma_{\eta}\pi\pi} \}\cdot (k+p+p')_{\mu} -
\frac{g_{\sigma_{K^0}K^+\pi^-}}{m^2_{\sigma_K} - (k-p)^2}\cdot (k - p +p')_{\mu}.
\end{array}
\label{22}
\end{equation}

Therefore, the form factors $f^{(\sigma)}_{1,2}$  are:
\begin{equation}
\begin{array}{l}
f^{(\sigma)}_1= m_K
 \left(\frac{2\sqrt{2}\cos\theta_S - \sin\theta_S}{m^2_{\sigma_{\eta'}} - (p+p')^2} \cdot g_{\sigma_{\eta'}\pi\pi}  -
\frac{2\sqrt{2}\sin\theta_S + \cos\theta_S}{m^2_{\sigma_{\eta}} - (p+p')^2} \cdot g_{\sigma_{\eta}\pi\pi} \right) -
f^{(\sigma)}_2 ,

\end{array}
\label{23}
\end{equation}
\begin{equation}
f^{(\sigma)}_2= \frac{m_K g_{\sigma_{K^0}K+\pi^-}}{m^2_{\sigma_K} - (k-p)^2}.
\label{24}
\end{equation}
At $p=0$, the values of $f^{(\sigma)}_{1,2}$ coincide with those
represented in (15). Consequently, our theory not only reproduces
the results of the algebra of currents and PCAC, but in addition,
permits to extrapolate them from the non-physical point
$(p+p')^2=\mu^2$ to the physical  region $(p+p')^2\ge 4\mu^2$,
where the experimental data exist. A dependence of the constants
$\theta_S, \; m_{\sigma_{\eta'}}, \;m_{\sigma_{\eta}}$ and their
coupling constants on a value of the parameter $\xi$ permits to
extract its magnitude from the data on $ f^{(\sigma)}_ {1,2}$. But
before, it is necessary to pick out from the data the possible
contributions of other scalar resonances and the contribution
produced by the spin-1 mesons.

The isosinglet scalar meson $f_0(980)$ does not enter into the
nonet of $\sigma_{0,...8}$ mesons. Its contribution into the
amplitude (13) is represented by matrix element $<\pi^+ \pi ^-|\sigma><\sigma|A_{\mu}^{(K)}|K^+>$.  The Adler
"self-consistency"  condition \cite{18} requires that the vertex
$f_0\pi\pi$ would have the form
$f_0(Q)\partial_{\mu}\pi(p)\partial_{\mu}\pi(p'),$
 or the form $f_0(Q)(Q^2-\mu^2)\pi(p)\pi(p')$.  Using the last form and taking into account that the axial current in the case of $f_0$ meson has the form
\begin{equation}
(A^{(f_0)}_{\mu})^{(K)}=\frac{2}{\sqrt{3}}(f_0\partial_{\mu}K^+ - K^+\partial_{\mu}f_0)
\label{25}
\end{equation}
we find the following expression for the additional contribution
of $f_0$ into the form factor $f_1(Q^2)$ :
\begin{equation}
\Delta f_1^{(f_0)}=4g_{f_0   \pi^+\pi^-}m_K(Q^2 - \mu^2)[\sqrt{3}m^2_{f_0}(m^2_{f_0} -
Q^2)]^{-1}.
\label{26}
\end{equation}
The constant $g_{f_0}$ could be determined using the data on the width  of $\Gamma_{f_0}(980)$. However, this width is very uncertain:
 $40\le \Gamma_{f_0} \le 100 $ MeV  \cite{17}.  At $\Gamma_{f_0} \approx 53 $ MeV
\begin{equation}     \Delta f^{(f_0)}_1 (4\mu^2)=0.108.
\label{27}
\end{equation}
This special choice will be cleared up later.
 But the others values of $\Gamma_{f_0}$ in the above-mentioned interval are not   forbidden too, because a value of $\Delta f^{(f_0)}_1(4\mu^2)$
changes in this interval from  0.094 to 0.1485 that leads to variation of $f_1(4\mu^2)<1\%$ while its experimental value is known with $1.4\%$ precision only.

\section{A contribution produced by the intermediate spin-1 mesons
to  the form factors $f_{1,2}$.}
In a theory where the vector fields $V^a_{\mu}$ are the divergences of the  antisymmetric tensor fields $V^a_{\mu\nu}$, the following relation takes place: \begin{equation}
\partial_{\mu}V^a_{\mu\nu}=MV^a_{\nu},
\label{28}
\end{equation}
where $M$ is a mass of the vector field $V^a_{\nu}$.

The part of effective lagrangian describing the  $V\pi\pi$ interaction is
\begin{equation}
L^{eff}(V\pi\pi)=- \frac{g^a}{M} V^a_{\mu\nu}f^{abc}D_{\mu}\pi^b D_{\nu}\pi^c.
\label{29}
\end{equation}
In the case of the  $\rho^{(3)}\pi\pi$ interaction we obtain the result:
\begin{equation}
  A(\rho^{(3)} (Q) \to \pi(p)\pi(p'))=
g_{\rho}\frac{Q^2}{M^2}\rho^{(3)}_{\mu}(Q^2) f^{3ab}\partial_{\mu}\pi^a(p) \pi^b(p'),
\label{30}
\end{equation}
 On the mass shell, this result coincides with the one following in the  gauge theory.
But at $Q^2<M^2$ it becomes smaller by $Q^2/M^2$  times than the result of the
gauge theory. This property will play an important role further.

The expression for the free propagator of the field $V_{\mu\nu}$ is \cite{10}, \cite{11}:
\begin{equation}
\begin{array}{cc}
< 0|T\{V_{\mu\nu}(x), V_{\rho\sigma}(y)\}|0>=
iM^{-2}\int \frac{d^4 k e^{-ik(x-y)}}{(2\pi)^4(M^2 - k^2 -i\epsilon)} \times    \\

[(g_{\mu\rho}g_{\nu\sigma} - g_{\nu\rho}g_{\mu\sigma})(M^2-k^2)+g_{\mu\rho}k_{\nu}k_{\sigma} -
g_{\mu\sigma}k_{\nu}k_{\rho} - g_{\nu\rho}k_{\mu}k_{\sigma}+
 g_{\nu\sigma}k_{\mu}k_{\rho}].
\end{array}
\end{equation}
The diagrams with the intermediate vector mesons bringing up a contribution into $K_{e4}$ decay
are represented by the matrix elements $<\pi ^-|A_{\mu}^{(K)}|K^{*0}><K^{*0}|\pi^+ K^->$ and
$<\pi^+ \pi ^-|\rho^0><\rho^0|A_{\mu}^{(K)}|K^+>$,
where $K^{*0}$ is the strange vector meson with a mass $m=896$ MeV and $\Gamma(K^{*0} \to K\pi)=48.7\pm 0.8$ MeV.

In the considered here problem, only the $W^{\pm}$ bosons play a role of the gauge ones. So that, the  covariant derivative in (29) has the form \cite{19}
\begin{equation}
D_{\mu}\hat \pi=\partial \hat \pi+i[\hat W_{\mu},\hat \pi]_- -[\hat W_{\mu}, \hat \sigma]_+.
\label{32}
\end{equation}
Retaining in the matrix $\hat \sigma$ only the non-zero vacuum expectations
of the $\sigma$ fields, namely, taking
\begin{equation}
\langle\hat \sigma\rangle=\mbox{diagonal} \frac{F_{\pi}}{\sqrt{2}}\{1, 1, (2R-1)\}
\label{33},
\end{equation}
we obtain for $D_{\mu}K^+$ in (29)
\begin{equation}
D_{\mu}K^+=\partial_{\mu}K^+ -\sqrt{2}F_{\pi}RW^+_{\mu}.
\label{34}
\end{equation}
Then, a calculation of the matrix elements $<\pi ^-|A_{\mu}^{(K)}|K^{*0}><K^{*0}|\pi^+ K^->$ and
$<\pi^+ \pi ^-|\rho^0><\rho^0|A_{\mu}^{(K)}|K^+>$  fulfilled basing on Eqs. (29), (31)
and (34) yields
\begin{equation}
\Delta f^{(K^*)}_1=-\frac{\sqrt{2}g^2_{K^*}RF_{\pi}m_K}{ M^2_{K^*}-(k-p)^2}\cdot\left(\frac{kp'-2pp'}{M^2_{K^*}}\right).
\label{35}
\end{equation}
And the  following contribution into $f_2$:
\begin{equation}
\Delta f^{(K^*)}_2=-\frac{\sqrt{2}g^2_{K^*}RF_{\pi}m_K}{M^2_{K^*}
-(k-p)^2} \cdot \left(\frac{kp'}{M^2_{K^*}}\right).
\label{36}
\end{equation}
\begin{equation}
\Delta f^{(\rho)}_2=-\frac{\sqrt{2}g^2_{\rho}RF_{\pi}m_K}{M^2_{\rho}-(p+p')^2} \cdot \left(\frac{kp+kp'}{M^2_{\rho}}\right)
\label{37}
\end{equation}
The well known existence of the $\rho$ meson dominance \cite{20}
implying inessention of the other intermediate states, permits us
to restrict the further  consideration  of spin-1  meson
contribution into the form factors $f_{1,2}$. The last multipliers
in the above formulae arising in a theory  with the vector hadrons
as the divergences of the antisymmetric tensor fields, diminish a
role of the vector hadrons in the low-energy meson processes.

In a theory including the mesons with the spin 0 and spin 1, the strange
axial-vector current, besides the part, presented in (19), contains additional part
\begin{equation}
A_{\mu}^{(K)} = \sqrt{2}F_{\pi}m_a a_{\mu}^{(K)} - \sqrt{2}F_{\pi}D_{\mu}K
\label{38} 
\end{equation}
Then an influence of the  intermediate axial-vector mesons is reduced to appearance
in the right-hand part of the relations (35)-(37) of the additional multiplier
\begin{equation}
\left(1 - \frac{\sqrt{2}F_{\pi}R m_a}{m_a^2 - (k-p-p')^2}\right).
\label{39}
\end{equation}
Limiting ourself by a consideration only the lightest axial-vector meson $K_1(1270)$
 contribution into $f_{1,2}$, we come to the results
\begin{equation}
\begin{array}{ccc}
\Delta f^{(K*)}_1(4\mu^2)= - 0.109706,\\
\Delta f^{(K*)}_2(4\mu^2)= - 0.226025, \\
\Delta f_2^{(\rho)}(4\mu^2)= -1.023669. 
\label{40}
\end{array}
\end{equation}

In our theory, the parts of $f_{1,2}$ originated by the intermediate scalar mesons  and
by the spin-1 mesons turn out to be of the same (negative) signs. Then, comparing
our results with the experimental values, we need to compare the absolute value of
$f^{theor}_{1,2}$ with $f^{exp}_{1,2}$.

The absolute value of the form factor $f^{theor}_2(4\mu^2)$ that is a sum of the
parts $f^{(\sigma)}_2(4\mu^2),  \Delta f_2^{(K*)}(4\mu^2),  \Delta f_2^{(\rho)}
(4\mu^2)$ is equal
\begin{equation}
|f_2^{theor}(4\mu^2)|=4.6837.
\label{41}
\end{equation}
This result practically coincides with the experimental value of $f_2^{exp}$
\begin{equation}
f_2^{exp}(4\mu^2)= 4.687(1\pm 0.024)
\label{42}
\end{equation}
obtained by extrapolation of the data to the point $(p+p')^2=4\mu^2$.
(We use for extrapolation the fitting formula $f(4\mu^2)=f(q^2)(1+\lambda \cdot q^2)^{-1}$, where $q^2=(s_{\pi}-4\mu^2)/ 4\mu^2$).

In a gauge theory, a magnitude of $f_2(4\mu^2)$ would be
\begin{equation}
|f_2(4\mu^2)|=2.7087    \qquad  \cite{21}.
\label{43}
\end{equation}
This value is  inadmissible. So
that, a treatment of the spin-1 hadrons as the gauge bosons is unacceptable,
Now, using the results (27), (35) and the numerical value of $f_1^{exp}$ extrapolated
to the  point $(p+p')^2=4\mu^2$
\begin{equation}
f_1^{exp}(4\mu^2)=5.812(1\pm 0.014)
\label{44}
\end{equation}
we find a value of $f_1^{(\sigma)}(4\mu^2)$
\begin{equation}
f_1^{(\sigma)}(4\mu^2)=f_1^{exp}(4\mu^2) - |f_1^{(K*)}(4\mu^2)+f_1^{(f_0)}
(4\mu^2)|=5.5943 \pm 0.081.
\label{45}
\end{equation}
The form  factor $|f_1^{(\sigma)}(4\mu^2)|$ in (23) acquires a very close value 
(namely, 5.5786) at
\begin{equation}
 \xi= - 0.216.
\label{46}
\end{equation}
  A knowledge of a value of $\xi$ permits us to fix the magnitudes of the constants
$\theta_S, \: g_{\sigma_{\eta'}\pi\pi}, \:  g_{\sigma_{\eta}\pi\pi}, \: m_{\sigma_{\eta'}}$ and $m_{\sigma_{\eta}}$. Using the numerical values  of these
constants presented in Appendix, one may check that in the soft-pion approximation
 (at $ p=0$) a value of $f_1^{(\sigma)}$ in  (23) coincides with a value of $f_1$
in (15).

\section{The masses and the widths of the isosinglet scalar mesons}
Using the formulae (9) and  the expressions for the coupling constants $g_{\sigma_{\eta'\pi\pi}}$ and $g_{\sigma_{\eta\pi\pi}}$ presented in Appendix,
we find:
\begin{equation}
m_{\sigma_{\eta'}}= 663 \: \mbox{MeV}, \quad \Gamma_{\sigma_{\eta'\pi\pi}}
= 767 \: \mbox{MeV},
\label{47}
\end{equation}
\begin{equation}
m_{\sigma_{\eta}} = 1369  \:\mbox{MeV}, \quad  \Gamma_{\sigma_{\eta\pi\pi}} =
 679 \:\mbox{MeV}.
\end{equation}
The above widths are calculated on the  mass shell. At the smaller energies their
dependence on $Q^2=(p+p')^2$ is given by the expression
\begin{equation}
\Gamma_{\sigma\pi\pi}(Q^2) =   \frac{3g^2_{\sigma\pi\pi}   \sqrt{1-4\mu^2/Q^2}}{32 \pi Q}
\label{49}
\end{equation}

\section{A probability of the decay $K^+\to \pi^+\pi^-e^+\nu$.}
Its probability is defined by the integral
\begin{equation}
w(K_{e4}) = \frac{G^2_FV^2_{us}m^5_K}{2^{14}\cdot 3 \cdot\pi^5}\cdot \\
\int \limits^1_{4\mu^2/m^2_K}[f^2_1(x)\cdot \Phi_1(x)+f^2_2(x)\cdot \Phi_2 (x)+
\frac{f^2_4}{m^4_K}\cdot \Phi_4(x)]dx
\label{50}
\end{equation}
where $x=Q^2/m^2_K$ and $f_4=\frac{\sqrt{2}m^3_K}{8\pi^2 F^3_{\pi}}$ \quad \cite{21}.
\begin{equation}
\Phi_1(x)=(1-8x+8x^3-x^4-12x^2\ln x) \left(1-\frac{4\mu^2}{m^2_K\cdot x}\right)^{1/2}
\end{equation}
\begin{equation}
\Phi_2(x)=\frac{1}{3}(1+72x^2-64x^3-9x^4+12(3x^2+4x^3)\ln x)\left(1-\frac{4\mu^2}{m^2_K\cdot x}\right)^{3/2}
\label{52}
\end{equation}
\begin{equation}
\begin{array}{l}
\Phi_4(x)=\frac{2}{3}m^4_K[(x+x^2)(1-8x+8x^3-x^4-12x^2\ln x)-  \\     \frac{4}{5}x
(1-x)^5]\left(1-\frac{4\mu^2}{m^2_K\cdot x}\right)^{3/2}
\end{array}
\label{53}
\end{equation}
Taking also into account the observed in  \cite{1}, \cite{22} a small growth of the
form factors in the region $4\mu^2\le Q^2\le m^2_K$ described by the linear fit
$$
f_j(q^2)= f_j(0) (1+\lambda_j q^2), \qquad  q^2=(Q^2-4\mu^2)/4\mu^2,
$$
that in terms of the variable $x$ transforms into
\begin{equation}
f_j(x)=f_j(4\mu^2)(1-\lambda_j+\lambda_j\frac{m^2_K}{4\mu^2}x)
\label{54}
\end{equation}
and using the result  \cite{1}
\begin{equation}
\lambda_{1, 2}=0.079 \pm 0.015
\label{55}
\end{equation}
we obtain
\begin{equation}
\int\limits_{4\mu^2/m^2_K}^1 f^2_1(x)\Phi_1(x)dx=f^2_1(4\mu^2)(4.164\pm 0.048)10^{-3},
\label{56}
\end{equation}
\begin{equation}
\int \limits_{4\mu^2/m^2_K}^1 f^2_2(x) \Phi_2(x)dx=f^2_2(4\mu^2)(0.8401\pm 0.0126)10^{-3}.
\label{57}
\end{equation}
\begin{equation}
\int\limits_{4\mu^2/m^2_K}^1 f^2_4 \Phi_4(x)dx=f^2_4\cdot 0.8213\cdot 10^{-5}.
\label{58}
\end{equation}
Then, using the fixed above magnitudes of $f_{1, 2}(4\mu^2)$, we come to the
result:
\begin{equation}
w(K_{e4})^{theor} = (3.2426 \pm 0.0781)10^3 s^{-1}.
\label{59}
\end{equation}
This value is in agreement with the observed result:
\begin{equation}
w(K_{e4})^{exp} = (3.302 \pm 0.078)10^3 s^{-1}
\label{60}
\end{equation}
\section{Conclusions}
The $K_{e4}$ decay turned out to be the perfectly suitable object
for a verification of the ideas on a nature of the particles
originating the low-energy processes with participation of the
pseudoscalar mesons.

Thus, a considerable exceeding of the observed value of the form factor $f_1$ its value calculated in the soft-pion approximation,  gets an explanation considering the structure of the diagrams in Fig.1, containing  the scalar mesons.

A fascinating idea, expressed by fifty years ago, that the spin-1
hadrons are the gauge bosons of chiral theory, remains recognized
up to now, in spite of the appearing from time to time doubt in
literature concerning its truth. The data on the form factors of
$K_{e4}$ decay permit to check how fair this idea.

We have found that in the theory, where  the spin-1 hadrons are considered as the gauge bosons, the
form factor $f_2(4\mu^2)$ turns out to be too small.

On the contrary, a theory treating the spin-1 hadrons as the divergences of the
antisymmetric tensor fields, gives the quite satisfactory results
for $f_2^{theor}(4\mu^2)$ and for $f_1^{theor}(4\mu^2)$ also.
\newpage
{\large  APPENDIX.}     \\
Here we present the formulae for the coupling constants of scalar mesons necessary
for calculation of the amplitudes
$$
g_{\sigma_{\eta'\pi\pi}}= - \frac{\Lambda^2}{\sqrt{3} F_{\pi}}\left\{\sqrt{2}[1+
\xi(2R+1)]\cos \theta_S+[1-4\xi (R-1)\sin \theta_S \right\},
$$
$$
g_{\sigma_{\eta\pi\pi}}= - \frac{\Lambda^2}{\sqrt{3} F_{\pi}}\left\{-\sqrt{2}[1+
\xi(2R+1)]\sin \theta_S+[1-4\xi(R-1)]\cos\theta_S \right\},
$$
$$
g_{\sigma_{K^0K^+\pi^-}}= - \frac{\Lambda^2 (2R-1)}{\sqrt{2} F_{\pi}}
$$
where $\Lambda^2=(m^2_K-\mu^2)/[(R-1)(2R-1)]$.
In the theory with the lagrangian (2), the following  relation takes place:
$$
\frac{g^2_{\sigma_{\eta'\pi\pi}}}{(m^2_{\sigma_{\eta'}}-\mu^2)^2}+\frac{g^2_{\sigma_
{\eta\pi\pi}}}{(m^2_{\sigma_{\eta}}-\mu^2)^2} = \frac{1}{F^2_{\pi}}
$$
The formulae for the other coupling constants, the reader can find
in [5] and in Appendix of the paper [24]. We present here also the
numerical values of the above constants and the masses of $\sigma$
mesons at $\xi= - 0.216$.
$$
\theta_S = 18.83728995 \:\hbox{degree}    \\
$$
$$
g_{\sigma_{\eta'\pi\pi}} = 4.335480173\: \mbox{GeV},\quad g_{\sigma_{\eta \pi\pi}} = 5.639539623 \:\mbox{GeV}\\
$$
$$
m_{\sigma_{\eta'}} =  0.663209405\:\mbox{GeV}, \quad
m_{\sigma_{\eta}} = 1.368980326\:\mbox{GeV}.
$$

\end{document}